\documentclass[12pt,preprint]{aastex}

\slugcomment{}
\shorttitle{Pulsar Wind Nebula in DA 495}
\shortauthors{Arzoumanian et al.}

\bibpunct[; ]{(}{)}{,}{a}{}{;}

\newcommand{\nh}{N_{\rm H}}

\newcommand{\asca}{{\em ASCA}}
\newcommand{\cxo}{{\em Chandra}}
\newcommand{\kothes}{KLR+08}

\begin{document}

\title{{\em Chandra\/} Confirmation of a Pulsar Wind Nebula in DA 495}

\author{Z. Arzoumanian\altaffilmark{1,2}}

\email{zaven@milkyway.gsfc.nasa.gov}

\author{S. Safi-Harb\altaffilmark{3}}

\author{T. L. Landecker\altaffilmark{4}}

\author{R. Kothes\altaffilmark{4,5}}

\and{}

\author{F. Camilo\altaffilmark{6}}

\altaffiltext{1}{CRESST and X-ray Astrophysics Laboratory, NASA-GSFC, 
Greenbelt, MD 20771.}

\altaffiltext{2}{Universities Space Research Association, Columbia, MD 21044.}

\altaffiltext{3}{Canada Research Chair. Department of Physics and Astronomy, University of
Manitoba, Winnipeg, MB, R3T 2N2, Canada.}

\altaffiltext{4}{National Research Council of Canada, Herzberg
Institute of Astrophysics, Dominion Radio Astrophysical Observatory,
Penticton, British Columbia, V2A 6J9, Canada.}

\altaffiltext{5}{Department of Physics and Astronomy, University of
Calgary, Calgary, AB, Canada.}

\altaffiltext{6}{Columbia Astrophysics Laboratory, Columbia
University, New York, NY 10027.}

\begin{abstract} As part of a multiwavelength study of the unusual
radio supernova remnant DA 495, we present observations made with
the {\em Chandra X-ray Observatory}. Imaging and spectroscopic
analysis confirms the previously detected X-ray source at the heart
of the annular radio nebula, establishing the radiative properties
of two key emission components: a soft unresolved source with a
blackbody temperature of 1 MK consistent with a neutron star,
surrounded by a nonthermal nebula 40$''$ in diameter exhibiting a
power-law spectrum with photon index $\Gamma = 1.6\pm0.3$, typical of a
pulsar wind nebula. The implied spin-down luminosity of the
neutron star, assuming a conversion efficiency to nebular flux
appropriate to Vela-like pulsars, is $\sim 10^{35}$ ergs~s$^{-1}$,
again typical of objects a few tens of kyr old. Morphologically, the
nebular flux is slightly enhanced along a direction, in projection
on the sky, independently demonstrated to be of significance in radio
polarization observations; we argue that this represents the
orientation of the pulsar spin axis. At smaller scales, a narrow
X-ray feature is seen extending out 5$''$ from the point source, a
distance consistent with the sizes of resolved wind termination
shocks around many Vela-like pulsars. 
Finally, we argue based on synchrotron lifetimes in
the estimated nebular magnetic field that DA 495 represents 
a rare pulsar wind nebula in which electromagnetic flux makes up a
significant part, together with particle flux, of the neutron star's
wind, and that this high magnetization factor may account for the
nebula's low luminosity. 
\end{abstract}


\section{Introduction}

With the {\em Chandra X-ray Observatory}'s first glimpse of the
Crab pulsar \citep{2000ApJ...536L..81W}, a new window was opened
onto the interactions of neutron stars with their surroundings.
Rotation-powered pulsars drive outflows of particles and
electromagnetic energy into the immediate environment; depending
primarily on the pulsar's age, this environment might simply be the
interstellar medium or, in the case of young, energetic pulsars, the
circumstellar debris of the supernova explosion that accompanied the
birth of 
the neutron star. The astrophysical value of these interactions lies
in their uniqueness as a probe of the content of pulsar winds, which
originate in the incompletely understood magnetospheres of neutron
stars, and of the physics of particle acceleration and shock mixing
of relativistic and non-relativistic fluids. In contrast to the
optical and radio observations that provided early insights into
these phenomena, X-ray studies, in particular with {\em Chandra}'s
unequalled angular resolution, sample the properties of energetic and short-lived
particles nearest the accelerating engines that produce them.

In the class of filled-center supernova remnants (SNRs), the bulk of the
radiation from a remnant can be traced back energetically to the
rapidly rotating neutron star at its core, with emission from the
shocked neutron star wind often defining a pulsar wind nebula (PWN). 
Shell-type remnants, by contrast, expend the energy of the
supernova explosion's blast wave. Evidence for the transfer of a
neutron star's rotational kinetic energy to its surroundings is
provided by the axisymmetric structures---arcs, rings, tori, and
jets---seen in high-resolution X-ray images of the environs of a number of pulsars.
Together with amorphous nonthermal emission also found around
energetic pulsars, these structures are brought to light through synchrotron
radiation by the pulsar's ultrarelativistic particle wind where it
is confined and perturbed by the surrounding medium; see, e.g.,
\citet{2006ARA&A..44...17G} for a review of PWNs. The morphological
and spectral properties of PWNs carry valuable information about the
physical mechanisms of energy transfer, the local conditions
(particle densities, flow velocities, magnetic field strengths,
etc.), and the nature of the particle-accelerating central engine.

\begin{figure}
\epsscale{0.95}
\plotone{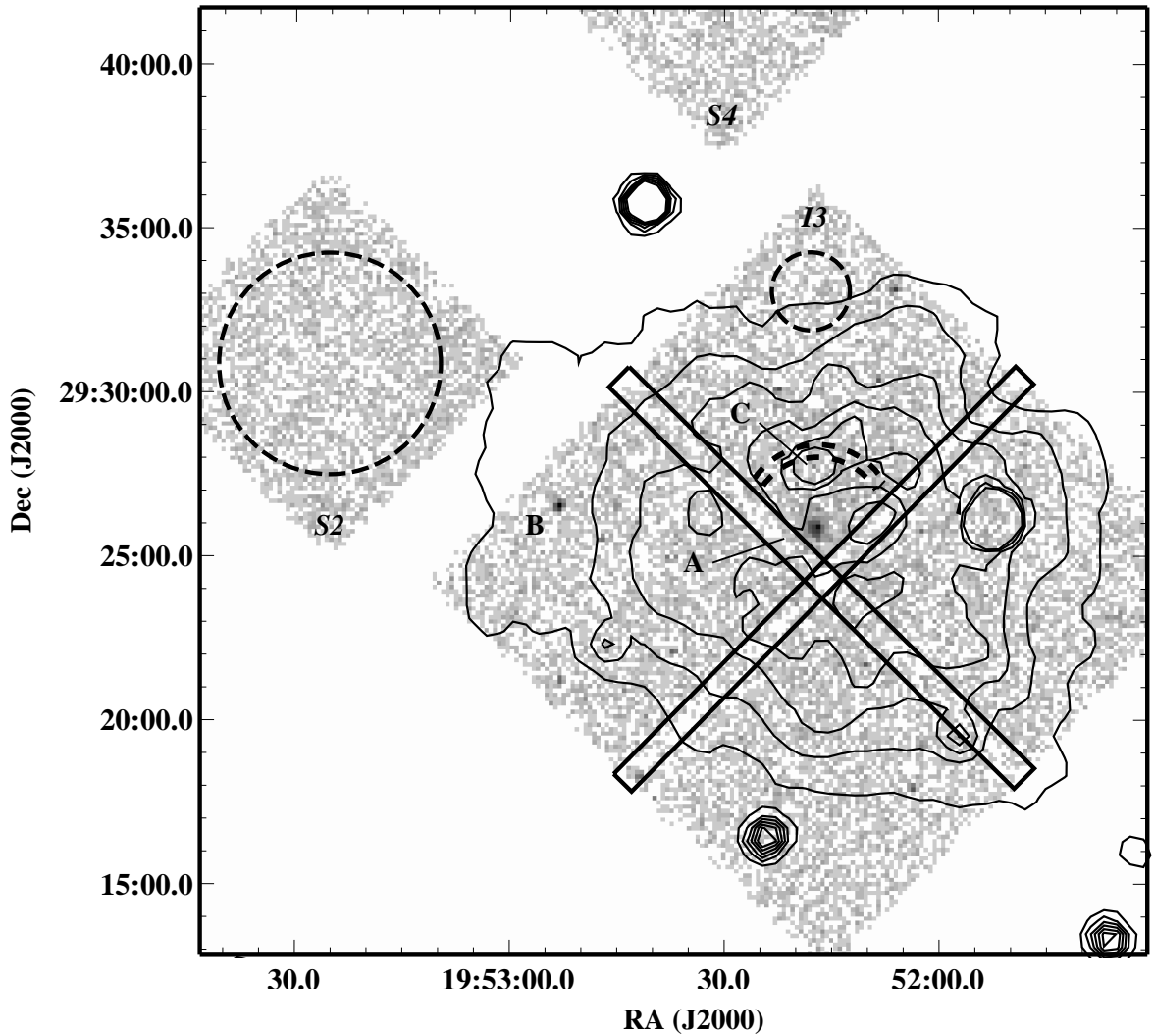}
\figcaption{\label{fig:radio}
\cxo-ACIS image of the DA 495 field in a 25 ks exposure:
X-ray counts in the 0.6--7 keV band, binned into $8''$-pixels, are
mapped on a logarithmic grayscale with a maximum of 218 counts.
Superposed contours show radio (1.4 GHz Canadian Galactic Plane
Survey [CGPS]; see ASLK04) brightness temperatures between 1 and 6 K
in 1 K steps. Three X-ray sources are labeled. Source A is the
proposed neutron star and wind nebula near the center of DA 495. Source B
is a Guide Star Catalog star previously detected by {\em ROSAT}
(1WGA J1952.8+2926). Source C, newly designated CXOU
J195217.3+292744, is apparently associated with an unresolved radio
enhancement to the north of source A. Event extraction regions used
for spectral background estimates are depicted by dashed circles and
arcs, while the solid rectangles show regions, affected by gaps
between ACIS-I chips, that were excluded from further analysis. }
\end{figure}

Although its identification as a pulsar wind nebula is now virtually
certain, the DA 495 radio nebula (G65.7+1.2; \citealp{1983AJ.....88.1810L})
is highly unusual: its morphology is distinctly
annular, with a radial profile that rises rapidly from a central
emission deficit (the ``hole'') then falls gradually with increasing distance
from the center; see Figure~\ref{fig:radio} for a contour map, and
\citet{klr+07} for a detailed discussion of the radio properties.
We described in an earlier paper \citep[hereafter,
ASLK04]{2004ApJ...610L.101A} X-ray observations of DA 495 that
revealed the central engine apparently powering the radio nebula, the {\em ROSAT}
source 1WGA J1952.2+2925, and provided some evidence for extended
X-ray emission consistent with a PWN. To confirm the existence of a
high-energy nebula, and because lingering doubts remained about a
possible ``thick shell'' thermal origin for the radio annulus
\citep{1989JApA...10..161V}, we set out to perform with \cxo\ an
X-ray observation sensitive to 
compact nonthermal structures as a means of 
strengthening the identification of DA 495 as a PWN, through
the detection in particular of jets, toroidal structures, or other
markers of large-scale axial symmetry, such as those seen around the
Crab, Vela, and similar pulsars.

\section{Observations}

The \cxo\ observation of DA 495 was carried out on 2002 December 9.
Anticipating the need for a large field of view---to search for
X-ray emission from the entire $\sim 20'$-diameter radio nebula---we
used the Advanced CCD Imaging Spectrometer (ACIS) instrument with
the putative pulsar, 1WGA J1952.2+2925 (see ASLK04), placed at the
aimpoint of the I3 CCD. The 4-chip ACIS-I array neatly captured most
of the radio extent of DA 495 within its $16.9'$ square field of
view. Two additional CCDs, the front-illuminated chips ACIS-S2 and
-S4 were also activated to provide a measure of the X-ray background
in fields adjacent to the radio nebula; 
ultimately, only S2, which
lies closer to the telescope's optical axis, was used.

X-ray photon events were acquired in timed-exposure (TE) readouts of
the ACIS CCDs, and telemetered to the ground in Very Faint mode. The
starting points for our analyses were event lists, aspect solutions,
and other standard data products resulting from the third
reprocessing (REPRO-III) of raw data at the \cxo\ X-ray Center. We
used the CIAO software package (version 3.3.0.1;
\citealp{2006SPIE.6270E..60F}) and associated calibration database
(CalDB version 3.2.3) for reduction of the ACIS data.

\section{Results}

Following standard data preparation procedures\footnote{Standard
procedures are described in CIAO software analysis guides and
``threads'' available at {\tt http://cxc.harvard.edu/ciao/guides/}.}
for filtering (e.g., rejection of events due to particle background
rather than cosmic X-rays), effective area and energy calibrations,
and mapping of detector to sky coordinates, final ACIS photon event
lists were used to cumulate images, spectra, and lightcurves for
further analysis. For an observation duration of 25,160 s, a total
filtered exposure of 24,843 s was acquired. We found that photon events
outside the 0.6--7 keV energy band were consistent with background
and provided little useful information; unless noted otherwise, all
of the following discussion refers to analysis and results in this
energy range.

\subsection{Image Analysis}

The map of X-ray counts on the sky for all CCDs active during the 25
ks \cxo\ observation, binned by a factor of 16, is shown in
Figure~\ref{fig:radio}, with radio contours superposed. The basic
result of earlier {\em ROSAT} and \asca\ observations (ASLK04), that an
X-ray source lies near the center of the DA 495 annulus, is readily
confirmed (source `A'). ASLK04 argued further that the \asca\ images
were consistent with a combination of soft emission from a compact source
and harder emission from a nebula barely resolved by the 
available $\gtrsim 1'$ angular resolution. 
The leap to {\em Chandra}'s arcsecond-level imaging upholds this
conclusion as well, demonstrating however (as discussed below) that
it was fortuitously reached: the extent of the diffuse X-ray
emission is too small to have been resolved by \asca. Instead, a
small number of unrelated nearby sources were blended by the \asca\
point-spread function, giving the appearance of extended nebulosity.

To allow for morphological analysis with the best available angular
resolution, we formed ACIS images with pixel randomization (used by
default in standard CIAO data reduction to avoid spatial
digitization effects) disabled; we also applied the sub-pixel event
localization algorithm described by \citet{2001ASPC..251..576M}. In
practice, we found that these techniques produced no discernible
change in the morphology of the diffuse X-ray emission, but yielded
a somewhat improved ($\sim 10$\%) concentration of counts onto the
point-like source from its immediate vicinity. The coordinates of
the unresolved source on the sky are (J2000) RA 19$^{\rm h}$52$^{\rm
m}$17\fs04, Dec +29\arcdeg25\arcmin52\farcs5, good to the nominal
\cxo\ astrometric uncertainty of $0.6''$.

\begin{figure}
\plotone{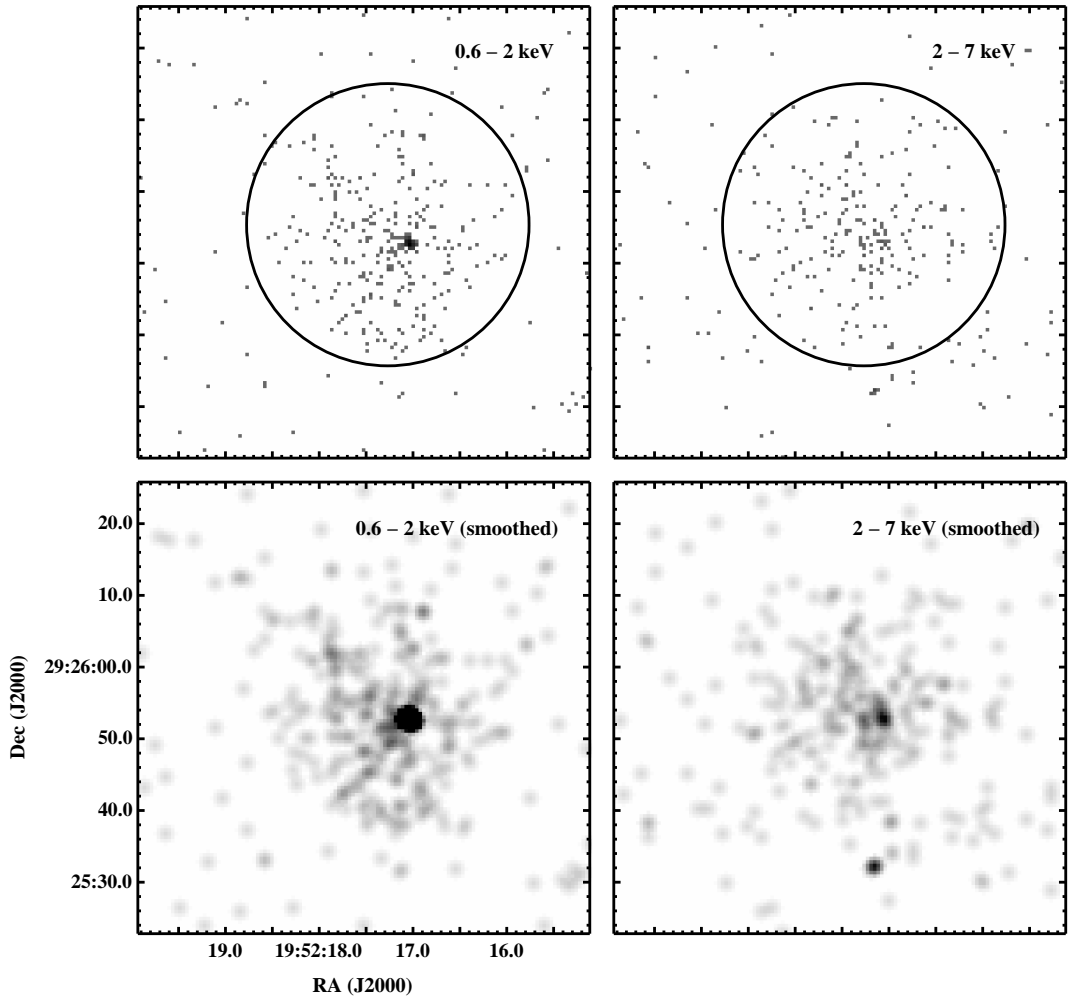}
\figcaption{\label{fig:twoenergies}
\cxo\ images of the central X-ray source in DA 495. 
{\em Upper panels:\/} Raw counts in two energy bands. The grayscale is
logarithmic, with a maximum of 50 counts for the central pixel of
the unresolved source in the 0.6--2 keV image. Solid circles depict
the extraction region for spectral analysis of the nebula; for
clarity, the $1''$-radius extraction region centered on the point
source is not shown. {\em Lower panels:\/} The same images
smoothed by convolving with a Gaussian of width $\sigma = 3$ pixels
($1.5''$). The grayscale is linear, truncated at 0.6 counts
pixel$^{-2}$.
} \end{figure}

\begin{figure}
\plotone{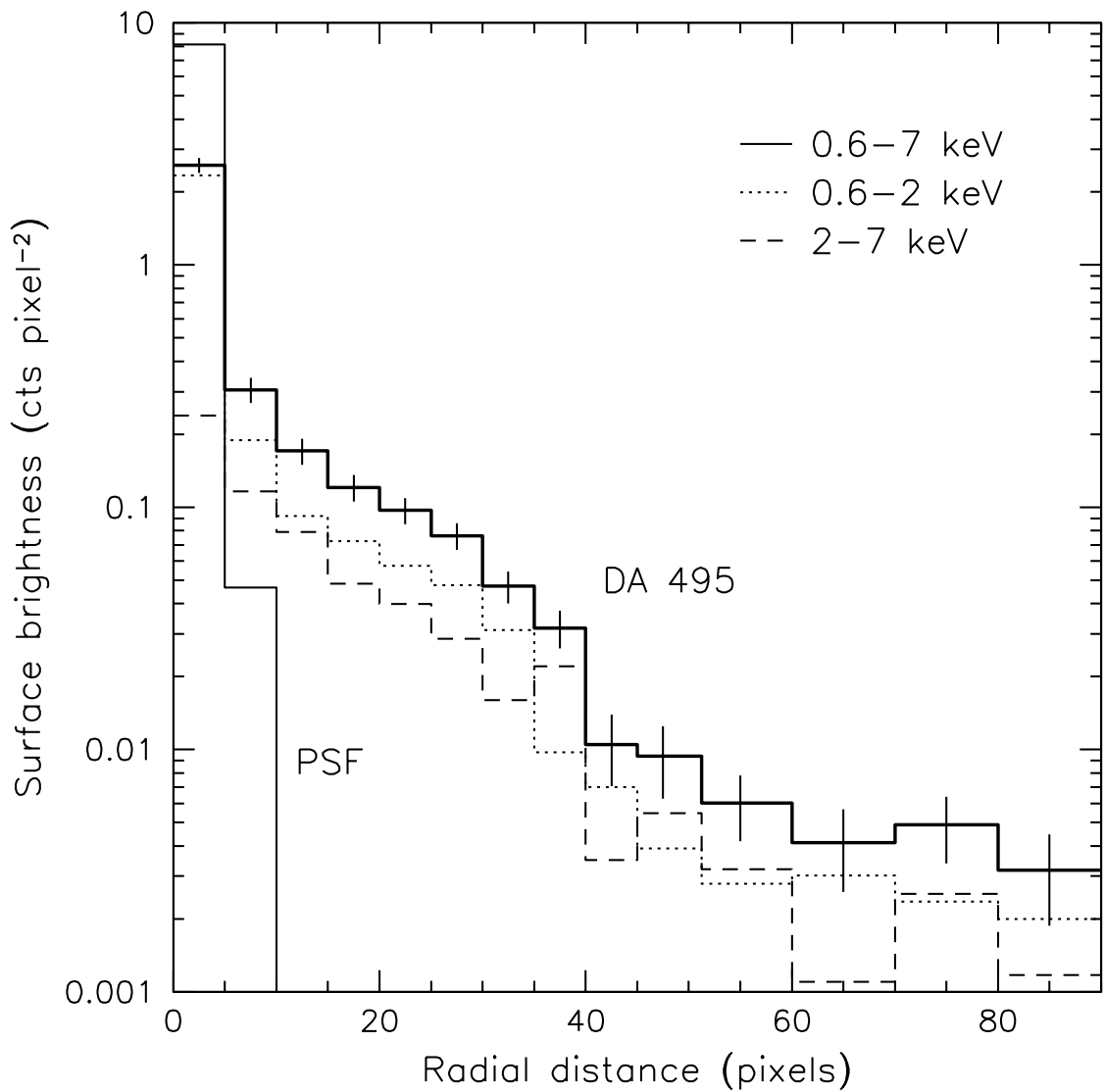}
\figcaption{\label{fig:psf}
Radial surface brightness profiles of the background-subtracted
X-ray emission from DA 495 in a series of annuli centered on the
unresolved source; the annuli have widths in radius of $2.5''$ 
within $25''$ (50 pixels) of the center, and $5''$ beyond. Photon
energy bands for the dotted, dashed, and solid lines are shown in
the legend. The thin solid line corresponds
to the radial profile of the \cxo\ point-spread function simulated
for a source at the same position as DA 495 on the ACIS-I detector
and with the same spectrum. By construction, the integrated area
under the two solid curves is the same.
}
\end{figure}

In the full-resolution ACIS-I image, Figure\ \ref{fig:twoenergies},
the X-ray nebulosity is revealed to be $\sim 40''$ in diameter. The
extended emission surrounds the point-like source but is not
centered on it. Instead, there is a strong asymmetry, roughly along
an east-west line, in both flux and radial extent, with most of the
diffuse flux lying to the east of the unresolved source. To quantify
this asymmetry, we sum the counts detected within the eastern and
western halves of a $20''$-radius circle centered on the point
source, neglecting the latter's flux by excluding the central
$1''$-radius circle. Of the 474 counts that define the X-ray nebula
and are contained within the larger circle, 314 lie to the east and
160 to the west of the north-south bisector. As shown in Figure
\ref{fig:twoenergies}, we choose an extraction region for spectral
analysis that is not centered on the point source, in order to
better match the asymmetric shape of the nebula. The
background-subtracted countrate within this extraction circle was 
$18.4\pm1.5$ cts~ks$^{-1}$. For the unresolved source, 175 counts
were detected in a $1''$-radius region centered midway between the
two brightest pixels, for a background-subtracted countrate of
$7.0\pm 1.3$ cts ks$^{-1}$; the same region was used to accumulate a
spectrum. Figure \ref{fig:twoenergies} displays the \cxo\
images in two energy bands, with and without smoothing; the softness
of the point source relative to the extended emission is evident.
Radial surface brightness profiles of the observed emission, in
comparison to the \cxo\ point-spread function, are shown in
Figure~\ref{fig:psf}.

\begin{figure}
\plotone{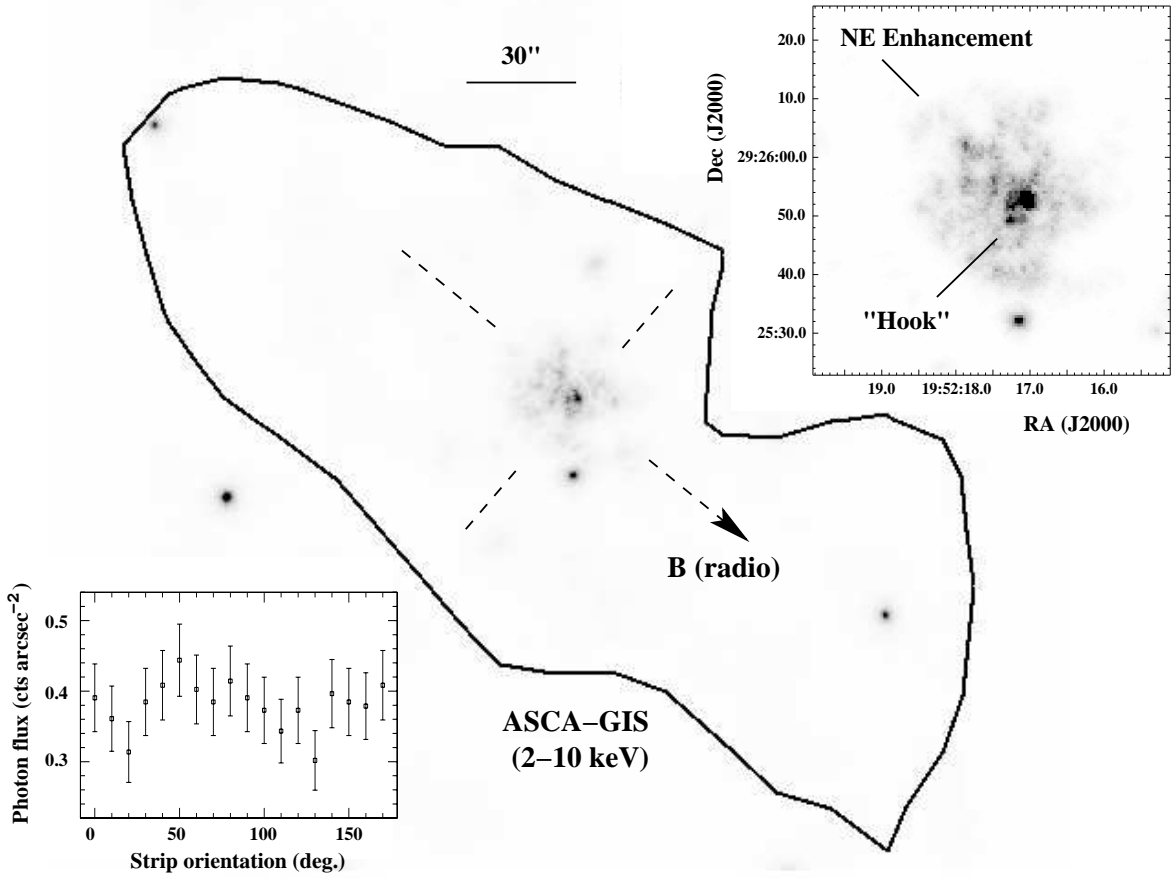}
\figcaption{\label{fig:jet}
Adaptively smoothed 2--8 keV \cxo\ image in the context
of an earlier \asca\ observation and the magnetic field
orientation derived from radio polarization measurements. 
For smoothing, a circular convolution kernel of varying size,
requiring a minimum of 5 counts within the kernel diameter, was
applied. The single
heavy contour depicts the 0.73 counts~pixel$^{-2}$ level in the smoothed
\asca-GIS image (Seq.\ 57054000, 45'' Gaussian smoothing; see 
ASLK04). \cxo\ resolves a handful of hard X-ray sources that are
responsible for the apparent extent of the low-resolution \asca\
morphology. The dashed lines show the polar ({\em arrow}) and
equatorial directions of the dipole field, on the plane of the sky,
inferred from the radio rotation measure 
(\kothes), consistent with the ``NE Enhancement'' in the X-ray PWN
and coincidentally in line with the \asca\ contour. {\em Inset,
upper right:\/} Close-up of the DA 495 PWN in the 0.6--7 keV band,
with adaptive smoothing as above.
The greyscale is linear, truncated at 1
count~pixel$^{-2}$. {\em Inset, lower left:\/} Summed counts within a
rectangular extraction region 5 pixels wide and 150 pixels long,
centered on the point source but excluding it, as a function of its
orientation angle north through east.
} \end{figure}

The low surface brightness of the X-ray nebulosity (0.36
cts~arcsec$^{-2}$) and the brevity of the observation conspire
to make difficult the identification of any well-defined structures
within the nebula. Smoothing of a raw counts image can increase the
significance of faint features if their extents are well-matched to
the smoothing scale, but the resulting images can also be misleading
if over-interpreted. Nevertheless, we experimented with Gaussian and adaptive
smoothing techniques to attempt to discern compact structures such
as those exemplified by the Crab nebula. Two candidate features
emerge. The first is an apparent arc of emission, resembling an
inverted hook, 10 pixels ($4.92''$) in extent and immediately
adjacent to the point source: it extends initially to the east and
then curves south (Fig.~\ref{fig:jet}). Whether this feature is truly curved is unclear,
but it is sufficiently bright in the unsmoothed image (61 counts in
a $15\times9$-pixel elliptical extraction region) relative to the local
background (an average of 21 counts in the same
elliptical region positioned at four nearby locations) that it
is statistically significant at the $4\sigma$ level. The apparent
curvature results from a deficit of counts immediately to the SE of
the point source, but statistical fluctuations could account for
this deficit.

The second structural detail suggested by the data, near the
detection limit for low surface-brightness features, is slightly
enhanced flux along a line (labeled ``NE Enhancement'' in Fig.\ \ref{fig:jet})
extending from the point source to the northeast. In Figure
\ref{fig:jet}, we plot the counts contained within a number of
identical rectangular strips, centered on the point source and
rotated at $10^\circ$ intervals. The resulting measurements are
found to be roughly consistent with a constant flux, but although
the deviations from the mean are not highly significant, the trend
is suggestive of an enhancement along the line at $50^\circ$ north
through east. Consistent with the
overall east-west asymmetry in the nebular flux, the enhanced
emission that may exist to the northeast is clearly nonexistent in
the opposite direction, to the southwest of the point source;
repeating the counting analysis shown in Figure \ref{fig:jet} with
a rectangular region that only samples the semicircular region to
the east of the point source (out to the same radius) results in a
statistically more significant peak at $50^\circ$.

While the evidence for this 
directionally enhanced flux
in the \cxo\ data
alone is 
meager, an independent measurement lends
support to make
it more compelling (cf.\ Fig.\ \ref{fig:jet}). In the work of
\citet[hereafter, KLR+08]{klr+07}, a multi-frequency radio
polarization study of DA 495 allows for the measurement of its
spatially-varying rotation measure. Combined with modeling of the
particle densities within the remnant, the orientation of the
intrinsic magnetic field can be derived. The result favors a dipolar
field oriented at $50^\circ$, identically to our NE enhancement, in
projection on the sky. Thus, two distinct results point to a
preferred direction and the possibility of axisymmetry in the
morphology of DA 495. 

We note with interest that the two diffuse features, the
hook and the NE enhancement, appear to lie in orthogonal directions, suggestive
of equatorial and polar structures seen in many PWNs. We discuss
this possibility further below.

In our analysis of \asca\ data
(ASLK04) we argued---without prior knowledge of the radio nor \cxo\
results---that a marginally resolved hard source was extended
along a NE-SW direction. Figure\ \ref{fig:jet} overlays the
innermost \asca-GIS smoothed-intensity contour onto the \cxo\ image.
Although the coincidence of the ASCA contour with the NE enhancement
and radio-derived magnetic axis is striking, {\em Chandra}'s
superior angular resolution reveals a collection of pointlike, hard
X-ray sources that collectively account for the elongated \asca\
feature, which is thus an artifact of the telescope's optical
limitations rather than a true morphological feature. 

Aside from the unresolved and diffuse X-ray emission believed to be
associated with DA 495, 48 additional sources are seen in
the ACIS-I field of view. 
Two of these lie within the confines of
the DA 495 radio nebula but their faintness and spectral properties
suggest that they are extragalactic background sources: 
\begin{description}
\item CXOU J195217.1+292532 is unresolved, faint (ten counts in a
$3\times3$-pixel area), hard, and lies $21''$ south of source A---it
is visible in the hard-band images in Figure~\ref{fig:twoenergies}. 
No counterpart at another wavelength has been cataloged.
\item CXOU J195217.4+292745 (Source C) consists of nine counts in a 
$3\times3$-pixel area, all between 2 and 8 keV, and is coincident
with an unresolved feature in radio images. Future high-resolution
radio maps will better distinguish between the nebular radio flux of
DA 495 and the contribution of source C, a likely background AGN.
\end{description}
A third notable source is the brightest in the field, CXOU
J195253.1+292633 (Source B). Astrometric coincidence suggests that
it is the X-ray counterpart of the star 2MASS J19525309+2926337.

\subsection{Spectral Analysis}

Spectral analysis was carried out for three source regions: the
point source at the heart of DA 495, the surrounding diffuse
emission within a $20''$-radius circle, and a larger annulus
encompassing the brightest parts of the radio nebula to search for
thermal X-ray emission associated with a ``thick shell'' supernova
remnant. Suitable background regions, in some cases more than one,
were identified for each source region; serendipitous sources were
excised and background spectra subtracted from accumulated source
spectra prior to fitting with the {\tt Sherpa} software. Results for
all spectral fits are summarized in Table~\ref{tab:spec}.

Following the results of \kothes, we adopt an assumed distance to DA
495 of 1 kpc when estimating luminosities of the various emission
components.

\subsubsection{Nebula and Point Source}

Point-source and nebular spectra were formed using the {\tt
psextract} and {\tt specextract} scripts, respectively, with
response matrices for the regions of interest generated using the
{\tt mkacisrmf} tool. For the point source, a circular extraction
region with a radius of 2 pixels ($\sim 1''$), centered midway between the two
brightest pixels, was defined to minimize contamination from the
nebula, and extracted events were grouped in energy so
that a minimum of 15 photons fell into each spectral bin. For the
diffuse emission, the extraction circle (Fig.\
\ref{fig:twoenergies}) was $40''$ in diameter and centered $4''$
to the northeast of the point source; events were grouped to a
minimum of 20 counts per energy bin. To allow for consistency
checks, two regions were
used to estimate the contribution of background emission to the
source spectra (see Fig.\ \ref{fig:radio}): an annular arc to the
north of the nebula with inner and outer radii $2.1'$ and $2.5'$
respectively, centered on the point source and extending out to the
chip gaps to its east and west; and a circular region (radius
$1.2'$) at the northern corner of the I3 chip, where there is
essentially no radio emission from the remnant. 

Events contained within the point source region reflect an evidently
soft-spectrum origin: just 4 counts were detected in the energy band
above 2.5 keV, consistent with the surface brightness of the
surrounding diffuse emission, so that the point source is cleanly
detected only at low energies (e.g., Fig.~\ref{fig:psf}). A blackbody
spectral model adequately describes this emission, with best-fit
parameter values that do not
depend, to within estimation uncertainties, on which background
region is used. 
In the diffuse emission, there were too few counts to 
allow for quantitative analysis of any features or sub-regions other
than the integrated nebula, which exhibits a hard tail consistent
with an absorbed power-law spectral model.

\begin{figure}
\epsscale{0.78}
\plotone{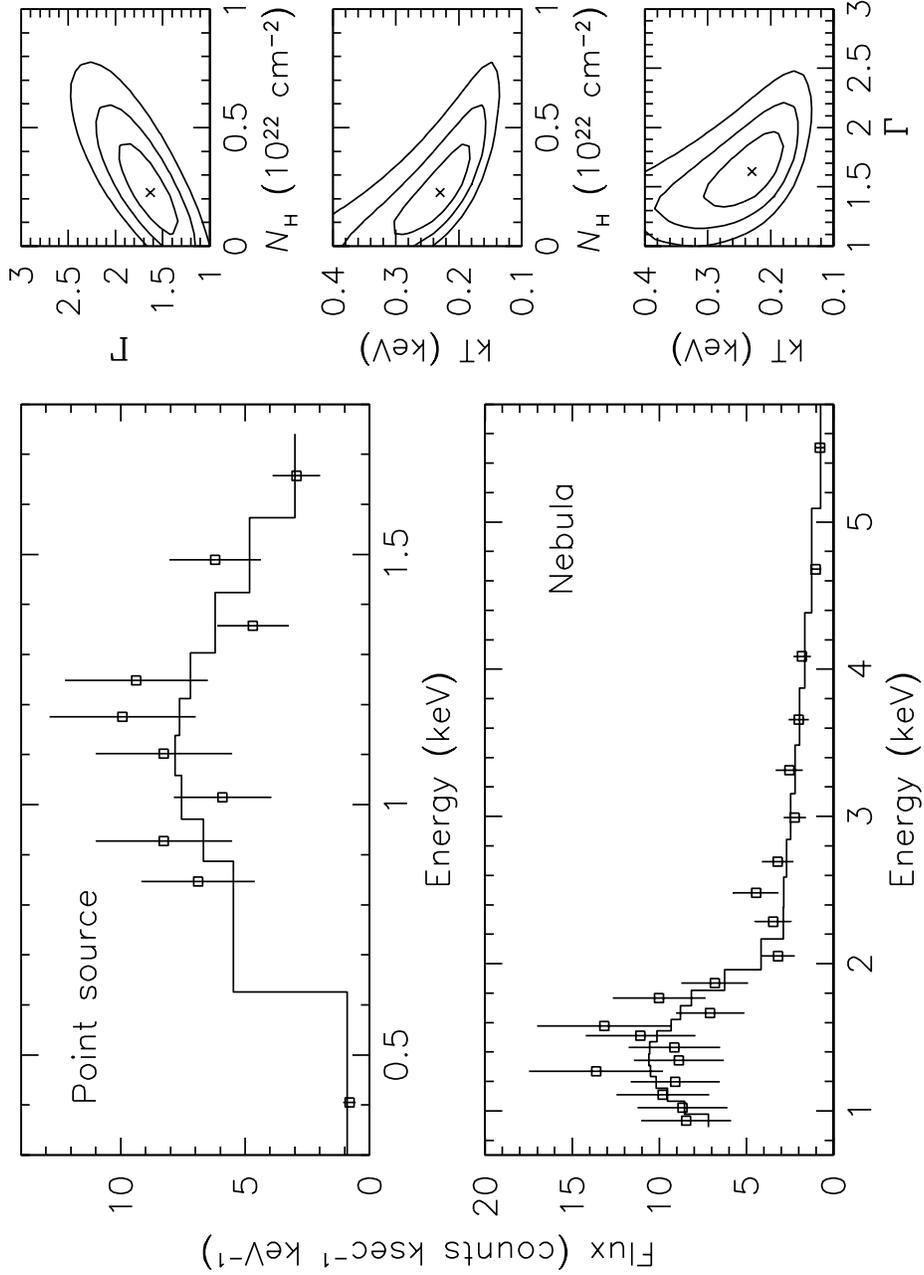}
\figcaption{\label{fig:spec}
Photon flux spectra of the unresolved X-ray source ({\em top left
panel}) and surrounding nebula ({\em bottom left}), fit
simultaneously to blackbody and power-law models respectively, with
a common neutral-hydrogen absorption ($\nh$) parameter. To the
right, projections of $\chi^2$ in three planes of interest are
plotted as contours representing 1, 2, and $3\sigma$ uncertainties
in the parameter values, where $kT$ is the blackbody temperature of
the point source and $\Gamma$ is the photon index of the nebula's
power-law spectrum.
} 
\end{figure}

Statistically, the most well-constrained spectral fits were those
made for the point-source (blackbody) and diffuse (power-law)
emissions simultaneously, with a common interstellar absorption
parameter, the neutral hydrogen column density $\nh$. We verified
that fits to each component separately produced consistent results,
finding however that poor statistics prevented measurement of $\nh$
in either fit alone---even in the simultaneous analysis, the best-fit
value $\nh = (0.23\pm0.17)\times 10^{22}$~cm$^{-2}$ amounts to an
upper limit at the $2\sigma$ level. Observed and model spectra
derived from the simultaneous analysis are shown in Figure
\ref{fig:spec}, together with confidence contours projected onto
two-dimensional slices of the full five-dimensional parameter grid.
Least squares minimization produced the best fit at a reduced
$\chi^2 \simeq 0.5$ for 27 degrees of freedom.

For the point source, power-law fits were formally acceptable
(reduced $\chi^2 \simeq 0.7$), but produced steep spectra (photon
indices $\Gamma \sim 3.5$) when $\nh$ was fixed at the nebular
value, and steeper still when $\nh$ was allowed to vary up to five
times the nebular value. We prefer instead thermal models: a
blackbody spectrum yields a best-fit temperature of
$0.23^{+0.06}_{-0.04}$ keV (uncertainties are $1\sigma$ throughout),
for a bolometric unabsorbed flux of $\sim 8\times
10^{-14}$~ergs~cm$^{-2}$~s$^{-1}$. The isotropic luminosity of the
source is then $L_X \sim 10^{31}\,(d / 1\;{\rm kpc})^2$
ergs~s$^{-1}$ for distance $d$, implying for a simple blackbody an
emitting area with a characteristic radius of $\sim 0.3\,(d /
1\;{\rm kpc})$ km. This size is consistent with the area of a 
magnetic polar cap, heated by bombardment of particles accelerated
in the magnetosphere, on the surface of a neutron star with a rotation
period of several tens of milliseconds (i.e., similar to the Vela
pulsar). Propagation effects
in stellar atmospheres, however, are believed to significantly
distort thermal spectra, making temperature and size estimates
unreliable, with pure blackbody fits tending to overestimate
temperatures by up to a factor of two depending on the magnetic
field strength and composition of the atmosphere. 

To investigate the possible impact of an atmosphere, we used the
{\tt XSPEC} NSA model
\citep{1995lns..conf...71P,1996A&A...315..141Z} to fit the
point-source spectrum to presumed emission from the entire stellar
surface for nonmagnetic atmospheres as well as $B = 10^{12}$ and
$10^{13}$ G conditions. For canonical neutron star mass (1.4
$M_\sun$) and radius (10 km) assumptions, these fits also provide a
distance estimate, as the emitting area seen by a distant observer
is fixed. The fits demonstrated that reasonable values of effective
surface temperature and distance (shown in Table \ref{tab:spec}; 
their implications are discussed in Sec.~\ref{sec:thermdisc})
could be inferred from the available data, but we found that the NSA
model parameters, effective temperature and overall normalization,
were highly covariant with one another in every case, with the
$\chi^2$ surface describing an extended, shallow minimum, precluding a
well-bounded measurement of either quantity by itself. The implied
luminosity, which depends on both temperature and the normalization,
is more reliably constrained. These ill-defined fits are likely the
result of poor statistics and the necessarily coarse binning of our
observed spectrum; we note, however, that the number of fitting
parameters was the same whether we used the NSA models or a simple
blackbody, where the latter produced much more robust parameter
estimations. Also unlike the blackbody model, the NSA fits all
required a hydrogen absorbing column $\nh$ some three times higher
than is supported by our preferred power-law fit to the nebular
emission surrounding the point source, as described below. When fit
simultaneously, $\nh$ remains high while the nebular spectrum
exhibits unmodeled excess flux at energies $\lesssim 1$ keV. Attempts 
to model this excess with additional thermal or nonthermal spectral
model components were inconclusive. 
A deeper observation will be needed to obtain reliable results from
neutron star spectral models incorporating atmospheric effects.

For the nebula, the best-fit photon index is $\Gamma = 1.63\pm0.27$,
for 0.5--10 keV and 2--10 keV unabsorbed fluxes of $2.8\times
10^{-13}$ and $1.8\times10^{-13}$~ergs~cm$^{-2}$~s$^{-1}$,
respectively. The implied 0.5--10 keV isotropic luminosity is $L_X
\simeq 3.3\times 10^{31}\,(d/1\;{\rm kpc})^2$ ergs~s$^{-1}$. These
results are largely in line with the \asca\ measurements reported by
ASLK04: the power-law photon index is virtually identical, but the
source luminosity derived from \asca\ data is a factor of 4 higher
than the \cxo\ value. This discrepancy 
is attributable to
contamination from the point source and other, unrelated, sources
that could not be distinguished from the PWN
in the low-angular-resolution \asca-GIS observation.
The newly measured $\nh$ is also consistent with the
\asca-derived value; at roughly one-sixth of the full Galactic
column density, it supports prior evidence of a small distance to DA
495.

\subsubsection{Constraints on nebular thermal emission}

The only significant diffuse X-ray emission in our \cxo\ image is the
nebula that lies within $\sim 40''$ of the point source, i.e.,
largely inside the radio ``hole.'' Fitting this emission to 
thermal models yields statistically acceptable results---with, e.g., 
a bremsstrahlung temperature of 10 keV---but these temperatures are
significantly higher than those typically seen in shock-heated
supernova remnant interiors ($\lesssim 3$ keV).
A thermal origin of the X-ray
nebula is therefore disfavored. More importantly, in a ``thick
shell'' interpretation of the radio emission as being thermal in
nature, one would expect significant spatial correspondence between
thermal X-ray emission and the bright radio annulus. No X-ray
emission in an annular or shell-like structure is seen. 

To quantitatively constrain X-ray flux 
coincident with the radio annulus, we defined an annular region with
inner and outer radii of $1.7'$ and $5.5'$ respectively, centered
on the radio hole, from which to extract source photons. Chip gaps
were excluded, and unresolved sources detected by the {\tt vpdetect}
tool were removed. To estimate the local background, we defined two
regions (as a check on the consistency of the results), one at the
northern corner of the I3 chip, and another taking up most of the S2
chip (cf.\ Fig.\ \ref{fig:radio}). The countrates in both of these
detector regions were degraded by $\simeq 15$\% due to mirror
vignetting, resulting in an overestimate of the source flux in a
background-subtracted analysis. For the purpose of setting a
conservative upper limit on the source flux, however, this is
acceptable. We find that the maximum allowable count rate in a
diffuse component within the radio-bright annulus is 0.048 cts
s$^{-1}$ ($3\sigma$) in the 0.5--5 keV band, for a surface
brightness $<15$ cts~arcmin$^{-2}$ in 25 ks; corrected for
vignetting, the upper limits on photon flux are $9.5\times
10^{-5}$ cts~cm$^{-2}$~s$^{-1}$ and $1.4\times 10^{-6}$
cts~cm$^{-2}$~arcmin$^{-2}$.

\subsection{Timing Analysis}

The photon time tagging resolution provided by the \cxo\ CCDs in
full imaging mode is 3.2 s, much too poor to allow for a periodicity
search for a typical young, rotation-powered pulsar. We did,
however, attempt to confirm the putative 10.9 s pulse period we
identified, with low statistical significance, in the \asca-GIS data
(ASLK04). In Fourier and Bayesian epoch-folding \citep{gl92b}
analyses, we detected no flux modulations at that, or any other,
pulse period $\gtrsim 10$ s, nor was there any evidence of flaring
or other variability in the point source and nebular fluxes.

\section{Discussion}
\label{sec:disc}

\subsection{Thermal properties of the point source and nebula}
\label{sec:thermdisc}

For the putative pulsar powering the DA 495 nebula, power-law
spectral fits are statistically acceptable but yield
uncomfortably large photon indices and significantly larger absorption
columns than are supported by the spectral fits to the nebula. 
For nominal fits to the NSA atmosphere models, the bolometric
luminosity for an assumed age may be compared with neutron star
cooling curves from theory and the blackbody temperatures and
luminosities of other pulsars with similar ages
\citep{2002ApJ...571L.143T}. \kothes\ estimate the age of DA 495 at
20 kyr. For this assumed age, the putative neutron star in DA 495 is
underluminous by a factor $\sim 5$ compared with expectations from
standard cooling models, $L_X \sim 10^{33}$ ergs~s$^{-1}$, assuming the
NSA fit results are reliable. 
The Vela pulsar, among others, is
similarly cooler and dimmer than expected \citep{2001ApJ...554L.189P}.
For the blackbody fits, the implied area and temperature are not
unreasonable for emission from a heated polar cap, but if the
observed flux is dominated by polar caps, 
any contribution from
cooling over the remaining surface would have to be significantly
(at least an order of magnitude) less luminous, further exacerbating
the discrepancy with cooling theory. It seems reasonable to assume, therefore,
that the observed thermal emission is dominated by whole-surface
cooling emission and that atmospheric effects are important. 

In the context of attempts to describe DA 495 as a ``thick shell''
SNR \citep{1989JApA...10..161V}, 
the absence of detectable diffuse thermal emission, either on large
scales or as in 3C58 where thermal and nonthermal emission are mixed
\citep{2007ApJ...654..267G}, suggests two possibilities. Either no
reverse shock has developed, and therefore the PWN is expanding into
cool surroudings (presumably ejecta), as in the Crab, or emissions
driven by the reverse shock have already dissipated. 
That a reverse shock, in the former scenario, has not yet produced
significant heating of ejecta nor interacted with the PWN in a $\sim
20$ kyr-old remnant would be somewhat surprising, but the absence of
a detectable limb-brightened shell surrounding DA 495 may be
significant---the existence of a forward shock is a prerequisite for
the formation of a reverse shock. 
The second case, in which
the hot ejecta that signal the passing of a reverse shock have
already cooled and dissipated, is difficult to reconcile with a
remnant age of just a few tens of kyr. Interaction with
a reverse shock would also be expected to significantly compress and distort a PWN
\citep{2001A&A...380..309V,2001ApJ...563..806B}, resulting in
asymmetries and, sometimes, spatial offsets between the radio and X-ray
emissions such as are found in the Vela and SNR G327.1$-$1.1
PWNs \citep{2006ARA&A..44...17G}.
The E-W asymmetry, relative to the central source, in the
morphology of the DA 495 X-ray nebula (a hint of which is also
evident in the radio morphology) could conceivably be
attributed to distortion from passage of a reverse shock, but there
is no measurable offset between the radio and X-ray nebulae,
both appearing to be tied to the current pulsar position to within,
at worst,
the $\sim 45''$ angular resolution of the available radio images.
On balance, it seems unlikely that interaction with a reverse shock
has occurred, so that DA 495 is not an evolved SNR.
The X-ray results therefore are at odds with a thermal origin for the
radio emission; \kothes\ provide further
arguments against the thick shell interpretation.

\subsection{Termination shock}

Pulsar wind nebulae are synchrotron structures that depend on the
perturbation of free-flowing particles at a standing shock. 
Of immediate interest in an observed nebula, then, is the
possibility of locating this wind-termination shock. Whether the wind
is confined by ram pressure induced by the pulsar's motion or simply
ambient pressure in a medium, the shock is associated with a
transition from an underluminous (free-flowing) zone to a brighter
region. The only feature in our \cxo\ image of DA 495 that
approximately matches this description is the ``hook'' near the
central point source: there is a hint of a deficit of X-ray flux
where the hook curves southward. If the deficit is significant, 
the hook may plausibly be associated with 
the standing shock where the pulsar wind is
equatorially confined. 
The extent of the hook, $\sim 4''$,
corresponds to a physical size of $r_{\rm hook} \sim 0.02\,d_1$ pc,
where $d_1 \equiv d/(1\;{\rm kpc})$ for distance $d$.

An alternative explanation for the hook is that it represents
post-shock emission from an equatorial outflow similar to the torus
around the Crab pulsar---for which the wind termination shock is more reliably
identified with an inner ring of emission---or an associated wisp,
as suggested for PSR B1509$-$58 by
\citet{2002ApJ...569..878G}\footnote{Such
wisps may arise from thermal instabilities due to
synchrotron cooling
\citep{1998MmSAI..69..883H} or compression of particle pairs by
magnetic reflections of heavy ions \citep{1994ApJ...435..230G}.}.
Indeed, the equatorial features seen in most X-ray PWNs are
thought to be extended toroidal structures a short distance downstream
of the termination shock, rather than ``inner ring'' markers of the
shocks themselves. In these cases, one-half of the observed radius
serves as a reasonable estimate of the true wind
termination shock radius $r_s$ (e.g., \citealp{2007ApJ...660.1413K}).
We adopt this interpretation for the DA 495 hook as well, so that
$r_s = r_{\rm hook}/2 \sim 0.01\,d_1$ pc, which is
consistent with (the low end of) the equatorial termination shock radii around several
Vela-like pulsars (see, e.g., Table 2 of \citealp{2007ApJ...660.1413K}). 
The combination of a $4''$-radius torus surrounding a $2''$-radius ring 
around an unresolved source would be difficult to
disentangle even with \cxo. We cannot rule out the possibility that
the termination shock is smaller still than we have assumed, that it
is unresolved and cannot be distinguished from the point source. The
Crab's inner ring is detected with surface brightness at least as
high as the downstream torus---in our dataset, the 4 counts above
2.5 keV within the point-source region represent a surface
brightness 50\% higher than that in the hook for the same energy
band, and Poisson fluctuations could allow an even higher unresolved
nonthermal flux. Nevertheless, our estimate of $r_s$ is at
worst a useful upper limit: no known PWNs
have estimated termination-shock radii smaller than 0.01 pc, and
arguments derived below from $r_s$ estimates generally require an
upper limit in any case.

Given, then, the tentatively identified shock radius, a
constraint on the spin-down luminosity of the pulsar is available
through pressure-balance,
\begin{equation}
\dot E = 4\pi c r_s^2 P_{\rm PWN}/f_\Omega,
\end{equation}
where $P_{\rm PWN}$ is the nebular pressure, and the factor
$f_\Omega$ accounts for a non-isotropic wind; 
a value $f_\Omega
\simeq 3/2$ corresponding to an equatorially-dominant wind is
typically assumed (see, e.g., \citealp{2002AstL...28..373B}).
Assuming equipartition between magnetic and particle pressures, 
$P_{\rm PWN}$ may be estimated from the
radio-derived average magnetic field, $B_{\rm neb} \simeq
1B_1$ mG (\kothes, and see Sec.~\ref{sync.sec}), where 
$B_1 \equiv B/(1\;{\rm mG})$,
such that $P_{\rm PWN} \simeq 8\,B_1^2\times 10^{-8}$ erg~cm$^{-3}$. The
resulting $\dot E$ is $2\,d_1^2\,B_1^2\times10^{37}$ ergs s$^{-1}$. 
An independent estimate of $\dot E$ is available, in
principle, by assuming an efficiency of conversion
from spin-down to nebular luminosity, $\eta \equiv L_X/\dot E$,
between $10^{-5}$ and $10^{-3}$, consistent with the PWNs seen
around Vela-like pulsars (i.e., those with characteristic ages of
10--30 kyr; \citealp{2007ApJ...660.1413K}). For our measured 0.5--8
keV nebular luminosity, $L_X = 2.9\times 10^{31}$ ergs~s$^{-1}$,
the implied $\dot E$ lies between $3\times 10^{34}$ and $3\times
10^{36}$ ergs~s$^{-1}$, well within the range of known Vela-like
pulsars, but a factor of 10--1000 below the value implied by the
putative termination shock radius, which is also consistent with the
same set of pulsars. Reconciliation of these $\dot E$ estimates
would require one of the following to hold true:
(1) the termination
shock is actually unresolvable ($\lesssim 1''$) and $\log\dot E\lesssim
36$; (2) $\dot E$ is indeed high ($\log\dot E\sim 37$) as required
by the hook, but the conversion efficiency $\eta$ to nebular flux is
only a few times $10^{-6}$; (3) the pulsar wind flows predominantly
into a solid angle much less than assumed by the factor $f_\Omega$, reducing the
pressure-derived $\dot E$ estimate by an order of magnitude or more;
(4) the nebular pressure estimate, which is based on
radio-wavelength measurements made at radii beyond the extent of the
observed X-ray nebula, is too high by an order of magnitude or more,
and $\log\dot E\sim 35$. Option (1) is possible, as argued above,
but would place DA 495 outside the company of Vela-like PWNs in its
termination shock radius,
while option (2) is difficult to entertain because a
$10^{37}$ ergs~s$^{-1}$ pulsar would seem likely to be detectable as
a radio or gamma-ray pulsar, or both, and would be significantly
younger than the estimated 20 kyr age of DA 495. Options (3) and
(4), or a combination of them, therefore appear most reasonable.
The assumption of equipartition in deriving $P_{\rm PWN}$, in
particular, may be suspect: in the PWN model of
\citet{1984ApJ...283..694K}, equipartition holds only at a single
post-shock distance that depends on the ratio of electromagnetic to
particle flux in the wind, with particle pressure dominating closer
in and magnetic pressure further out. Typical equipartition radii
are 5--$20 r_s$. Because the radio nebula is brightest at $\gtrsim
30 r_s$, the pressure inferred from the radio-derived magnetic field
may significantly overestimate the confining pressure at the wind
termination shock. For pressures that correspond to equipartition
field strengths comparable to those of PWNs around other Vela-like
pulsars, $B_1 \simeq 0.1$, the two methods of estimating $\dot E$
are brought in line.

\subsection{Morphology and connection to radio properties}

Lending further credence to the identification of DA
495 as a PWN is the fact that the combination of a soft point source
surrounded by a hard nebula is not uncommon among pulsars that are
tens of kyr old. In particular, the Vela pulsar, PSR B1706$-$44, and
G189.22+2.90 (IC443) all share this property
(\citealp{2001ApJ...554L.189P}, \citealp{2002ApJ...567L.125G}, and
\citealp{2006ApJ...648.1037G}, respectively), where Vela and
B1706$-$44 inflate wind nebulae confined by the ambient medium while
IC443's supersonic motion produces a ram-pressure confined nebula.
Intriguingly, sensitive radio imaging and polarimetry of the Vela
supernova remnant has shown \citep{2003MNRAS.343..116D} that the
pulsar and PWN lie within a radio emission depression straddled by a
pair of symmetric polarized lobes, both of which are also features
of DA 495.

That the pulsar wind in DA 495 is confined by ambient rather than
ram pressure is supported by two observations: (1) the X-ray PWN
morphology does not resemble a bow shock or narrow ``trail'' of
emission (cf.\ PSR B1957+20; \citealp{2003Sci...299.1372S}), and (2)
despite the spatial offset of the point source from the center of
the radio annulus, a low velocity---at most a few tens of
km~s$^{-1}$---is suggested by the anchoring of
the polarized radio emission (which is due to long-lived particles
injected soon after the pulsar was born) at the current position of
the pulsar (\kothes). The radio hole, then, may plausibly be
attributed to an absence of synchrotron-emitting particles in a
region similar to the so-called optical ``bays'' seen in the Crab
nebula \citep{1992ApJ...399..599F}. Regrettably, the spacecraft roll
angle during the observation was such that the coordinates of the
radio hole largely fell upon one of the chip gaps in the ACIS-I
array (Fig.~\ref{fig:radio}), resulting in poor sensitivity to any
X-rays from the hole region.

Virtually all known PWNs are axisymmetric in their morphologies, and
many---e.g., 3C58 \citep{2004ApJ...616..403S} and the
PWN in the G54.1+0.3 SNR \citep{2002ApJ...568L..49L}---
are notably elongated in X-rays along their apparent pulsar spin axis directions. 
The only feature in the DA 495 X-ray PWN that hints at axisymmetry is the
putative NE enhancement; its existence depends on just a
dozen or so detected photons in our \cxo\ image, but the symmetry axis 
that it suggests is supported by an independent radio measurement.
Whether the NE enhancement in DA 495 represents an elongation
of the PWN's faint outer reaches or is simply a brightened region is difficult to
establish conclusively from the available data. If its
orientation does correspond to the spin axis of the neutron star,
the feature may plausibly be explained by increased synchrotron lifetimes
in a weak axial field relative to the equatorial direction, 
or perhaps a variation on the proposed process
\citep{2002MNRAS.329L..34L,2004A&A...421.1063D} that collimates
axial outflows into the narrow relativistic jets that appear to
emanate from the rotational poles of the Vela pulsar, B1706$-$44,
and many others.

Aside from orientation, the radio model of \kothes\ provides 
further evidence in support of the NE enhancement in DA 495:
the radio polarization-derived inclination of the magnetic field
vector to the line of sight, $\sim 55^\circ$, shows the field
pointing away from us to the SW, so that the \cxo\ brightening to
the NE (and perhaps much of the overall E-W asymmetry in X-ray flux)
is consistent with Doppler boosting, especially given the
relativistic flow speeds implied by the synchrotron cooling time
(see below). If the ``hook'' to the east and
south represents a portion of the wind termination shock or an
equatorial torus, then the
DA 495 X-ray structure would be reminiscent of the PWN in G54.1+0.3
and other objects (e.g.,
\citealp{2004ApJ...601..479N}) in that both the equatorial and polar
(jets, in most cases) structures, ostensibly orthogonal to one
another, are brightened on the same side of the pulsar, a situation
difficult to reconcile entirely with Doppler boosting. 

Comparison of the radio and X-ray spectra is also instructive.
\kothes\ find that the broadband radio spectrum of DA 495 shows a
break near 1.3 GHz; the measured spectral slopes below and above the
break are $\alpha_{\rm Rlo} = -0.45 \pm 0.2$ and $\alpha_{\rm Rhi} =
-0.87 \pm 0.1$ respectively (defined such that flux $S$ at frequency $\nu$ 
is expressed as $S \propto \nu^\alpha$). 
\kothes\ weigh the available evidence 
bearing on the origin of the break, whether it is due to synchrotron
cooling or is intrinsic to the spectrum of radiating particles
injected by the pulsar. A key discriminant is the spectrum of the X-ray
PWN, which suggests through the following reasoning that the break
is in fact due to cooling and not a consequence of the injected spectrum
(see \kothes\ for additional arguments). The radiative lifetimes of
synchrotron-emitting particles are such that the bulk of the
observed radio emission must arise from particles that were injected
into the nebula soon after the pulsar was
born, when spin-down luminosity was at its peak. 
In contrast, the X-ray emission is dominated by
particles that have only recently been injected (see below). Thus,
if the origin of the break lies in synchrotron cooling, if the
injected particle spectrum is unbroken between the radio and
X-ray bands, and finally if the injected spectrum has not
changed substantially since the pulsar was born (despite
declining in particle luminosity) then one expects that the 
X-ray spectral slope $\alpha_X \equiv 1-\Gamma = -0.63\pm 0.27$
should lie somewhere between the two radio spectral slopes, and
perhaps closer to $\alpha_{\rm
Rlo}$. This is indeed observed, albeit with 
large combined uncertainties. The general agreement in
spectral slopes suggests that the conditions above are not violated.
For example, the existence of an injected break at an intermediate
frequency between radio and X-ray can be discounted at the $2\sigma$
level if it is to resemble those with spectral slope differences
$\Delta\alpha > 0.5$ seen at tens of GHz in three filled-center SNRs
(G29.7$-$0.3, 3C58, and G16.7+0.1; see \kothes). 
Similarly, attribution of the observed break at 1.3 GHz to the
injected spectrum is formally allowed but somewhat
contrived: it would require that $\alpha_X$, $\alpha_{\rm Rhi}$,
and $|\alpha_{\rm Rhi} - \alpha_{\rm Rlo}|$ {\em all\/} take on
values at or beyond their $1\sigma$ uncertainties so that the X-ray slope
is at least as steep as the high-frequency radio
slope, and that the magnitude of the 1.3 GHz break is greater than
0.5, for consistency with other injected breaks in the radio band.

\subsection{\label{sync.sec}Wind magnetization}

Based on the conclusion that the 1.3 GHz spectral break is due
to radiation losses, \kothes\ compare the energetics of DA 495 with
those of other PWNs and infer a nebular field strength
of $B_{\rm neb} \sim 1.3$ mG, a value that exceeds the Crab Nebula's magnetic field
by a factor of four. In addition to the radio probes, such as rotation
measure, of $B_{\rm neb}$, an independent determination is possible in
principle through steepening of the X-ray emission spectrum with
increasing radial distance from the pulsar, but 
because of the short \cxo\ exposure and low surface brightness of 
DA 495 in X-rays, a significant measurement was not possible.
The vastly different sizes of the X-ray and radio nebulae suggest,
however, that synchrotron cooling is very efficient, consistent with
the high field derived by \kothes---the Crab Nebula, for
instance, has a radio-to-X-ray size ratio of roughly 3.3, whereas
for DA 495 the ratio is $>25$. Because there is substantial
uncertainty in the value of $B_{\rm neb}$, however, we assume a
value of 1 mG and carry through scalings with
$B_1 \equiv B/(1\;{\rm mG})$
when estimating derived quantities.

The synchrotron lifetime of particles emitting at energy
$\epsilon$ keV in a magnetic field $B$ $\mu$G is $\tau_{\rm
synch} = 39 B^{-3/2} \epsilon^{-1/2}$ kyr.
For a 1 mG field
and emission at $\epsilon = 1.5$ keV, the radiating
lifetime is $\tau_{\rm synch} = 1\,B_1^{-3/2}$ yr. 
Thus, the DA 495 PWN
imaged with \cxo\ is dominated by particles roughly one 
year old. The velocity necessary to
traverse the $\gtrsim 20''$, or $0.1\,d_1$ pc,
radius of the nebula within the radiating lifetime is $0.33\,B_1^{3/2}c$. 
\citet{2002ApJ...568L..49L} find a velocity a substantial
fraction of $c$ for the
G54.1+0.3 PWN in the flow immediately downstream of the
termination shock; 
at larger post-shock distances, a velocity of $\sim 0.01 c$ has been estimated
for PSR B1046$-$58 \citep{2006ApJ...652..569G}.
In the case of DA 495, 
the relativistic flow must extend to a great distance post-shock,
given that the hook feature provides a reliable upper
bound on the termination shock radius $r_s$ at roughly one-tenth the PWN
radius. 

Relativistic bulk velocity at a large downstream
distance has been inferred for a linear feature in the N157B PWN
\citep{2006ApJ...651..237C}, but is thought to arise for that
object from strong
pressure confinement due to a nearby dense cloud. 
In the absence of such external influence, the
work of \citet{1984ApJ...283..694K} suggests that relativistic bulk
motion at large radii is only possible for values of the
``magnetization parameter,'' the ratio of electromagnetic to
particle energy flux in the pulsar wind, of $\sigma \simeq 0.3$
which, we argue, applies to DA 495. By
contrast, for G54.1+0.3, $\sigma = 0.06$, while for the Crab it is
smaller still, $\sigma = 0.003$. 
In fact, most estimates of $\sigma$
to date have found that the observed PWNs are driven by strongly
particle-dominated winds (e.g., \citealp{2007ApJ...662..988P}). 
A notable exception is the Vela X-ray PWN, for which
\citet{2001ApJ...556..380H} argue that $\sigma$ is nearly unity,
based on the \cxo-resolved termination shock and an earlier estimate
of the nebular magnetic field. \citet{2003ApJ...593.1013S} reach a
similar conclusion from modeling of the broadband spectrum. Thus, DA
495 provides, potentially, a second example of a pulsar wind in
which Poynting and particle flux appear 
to be comparable.
This conclusion is based on an inferred
nebular magnetic field and the observed size of the X-ray PWN;
it is worth keeping in mind that spin-down luminosity
considerations suggest that the nebular pressure, and thus magnetic
field strength, in the X-ray nebula may be
smaller than implied by the radio $B$-field measurement made at
larger distances from the neutron star, underscoring the need for a direct measurement of
synchrotron lifetimes through future, high-sensitivity spectroscopy
of the X-ray nebula to constrain the burn-off rate via radial
steepening of the spectrum.

Importantly, \citet{1984ApJ...283..694K}
show that, naturally, particle dominated winds are necessary for
efficient conversion of wind luminosity to synchrotron luminosity.
\citet{2004HEAD....8.0807G} has pointed out that pulsars
with spin-down luminosity less than $\simeq 4\times 10^{36}$
ergs~s$^{-1}$ host substantially dimmer PWNs than their more luminous
counterparts. It is tempting to speculate that the magnetization
fraction $\sigma$ varies either with spin-down luminosity or
with time such that older, less luminous pulsars produce dim
nebulae because Poynting flux contributes substantially to their winds. Current
understanding of magnetospheric physics for rotation-powered pulsars
argues for a strongly field-dominated wind at the light-cylinder,
$\sigma_{\rm LC}\sim 10^4$. 
The much smaller measured values 
$\sigma \lesssim 0.01$ in PWNs imply conversion to a
strongly particle-dominated wind by the time the flow reaches the
termination shock, at least for the youngest and most energetic
pulsars. Candidate mechanisms for effecting this conversion have
been proposed \citep{2002ApJ...566..336C,2007A&A...466..301C}, but
confrontations with observation have not yet been possible. If
future PWN studies confirm present suggestions that the winds of
Vela-like pulsars emerge from confinement with $\sigma$ of order
unity, the implied reduction in the efficacy of the conversion
mechanism for older, less luminous pulsars will be an important test
of these models.

\section{Conclusions}

The soft spectrum of the unresolved source near the center of the DA
495 radio nebula, its implied blackbody temperature and luminosity,
and its lack of variability on timescales of a few seconds through
the 25 ks duration of the observation are highly suggestive of
emission from the surface of an isolated neutron star. Taken
together with the properties of the surrounding extended, nonthermal
emission,
no doubt remains that DA 495 is a pulsar wind nebula,
driven by the rotation-powered magnetosphere of a neutron star.
Its unusual radio
characteristics can be explained as the result of aging in the absence of
any evident interaction with the supernova explosion's reverse
shock, such that DA 495 presents a valuable example of a pristine
PWN at an advanced age.

\acknowledgements
We thank the anonymous referee for a rigorous review that resulted
in a much-improved manuscript.
Support for this work was provided by the National Aeronautics and
Space Administration through \cxo\ Award Number GO3-4092A issued by
the \cxo\ X-ray Observatory Center, which is operated by the
Smithsonian Astrophysical Observatory for and on behalf of the
National Aeronautics and Space Administration under contract
NAS8-03060. SSH is supported by the Natural Sciences and Engineering
Research Council (NSERC) of Canada and the Canada Research Chair
program. The CGPS is a Canadian project with international
partners and is supported by the Natural Sciences and Engineering
Research Council (NSERC).

\bibliographystyle{apj}
\bibliography{apj-jour,myapj-jour,psrrefs,./ms}

\begin{thebibliography}{40}
\expandafter\ifx\csname natexlab\endcsname\relax\def\natexlab#1{#1}\fi

\bibitem[{{Arzoumanian} {et~al.}(2004){Arzoumanian}, {Safi-Harb}, {Landecker},
  \& {Kothes}}]{2004ApJ...610L.101A}
{Arzoumanian}, Z., {Safi-Harb}, S., {Landecker}, T.~L., \& {Kothes}, R. 2004,
  \apjl, 610, L101 (ASLK04)

\bibitem[{{Blondin} {et~al.}(2001){Blondin}, {Chevalier}, \&
  {Frierson}}]{2001ApJ...563..806B}
{Blondin}, J.~M., {Chevalier}, R.~A., \& {Frierson}, D.~M. 2001, \apj, 563, 806

\bibitem[{{Bogovalov} \& {Khangoulyan}(2002)}]{2002AstL...28..373B}
{Bogovalov}, S.~V. \& {Khangoulyan}, D.~V. 2002, Astronomy Letters, 28, 373

\bibitem[{{Chen} {et~al.}(2006){Chen}, {Wang}, {Gotthelf}, {Jiang}, {Chu}, \&
  {Gruendl}}]{2006ApJ...651..237C}
{Chen}, Y., {Wang}, Q.~D., {Gotthelf}, E.~V., {Jiang}, B., {Chu}, Y.-H., \&
  {Gruendl}, R. 2006, \apj, 651, 237

\bibitem[{{Contopoulos}(2007)}]{2007A&A...466..301C}
{Contopoulos}, I. 2007, \aap, 466, 301

\bibitem[{{Contopoulos} \& {Kazanas}(2002)}]{2002ApJ...566..336C}
{Contopoulos}, I. \& {Kazanas}, D. 2002, \apj, 566, 336

\bibitem[{{Del Zanna} {et~al.}(2004){Del Zanna}, {Amato}, \&
  {Bucciantini}}]{2004A&A...421.1063D}
{Del Zanna}, L., {Amato}, E., \& {Bucciantini}, N. 2004, \aap, 421, 1063

\bibitem[{{Dodson} {et~al.}(2003){Dodson}, {Lewis}, {McConnell}, \&
  {Deshpande}}]{2003MNRAS.343..116D}
{Dodson}, R., {Lewis}, D., {McConnell}, D., \& {Deshpande}, A.~A. 2003, \mnras,
  343, 116

\bibitem[{{Fesen} {et~al.}(1992){Fesen}, {Martin}, \&
  {Shull}}]{1992ApJ...399..599F}
{Fesen}, R.~A., {Martin}, C.~L., \& {Shull}, J.~M. 1992, \apj, 399, 599

\bibitem[{{Fruscione} {et~al.}(2006){Fruscione}, {McDowell}, {Allen},
  {Brickhouse}, {Burke}, {Davis}, {Durham}, {Elvis}, {Galle}, {Harris},
  {Huenemoerder}, {Houck}, {Ishibashi}, {Karovska}, {Nicastro}, {Noble},
  {Nowak}, {Primini}, {Siemiginowska}, {Smith}, \&
  {Wise}}]{2006SPIE.6270E..60F}
{Fruscione}, A.,
  et~al.\ 2006, in Proc.\ of the SPIE 6270,
  Observatory Operations: Strategies, Processes, and Systems, ed.\ D.~R. Silva,
  R.~E. Doxsey (Bellingham: SPIE), 62701V

\bibitem[{{Gaensler} {et~al.}(2002){Gaensler}, {Arons}, {Kaspi}, {Pivovaroff},
  {Kawai}, \& {Tamura}}]{2002ApJ...569..878G}
{Gaensler}, B.~M., {Arons}, J., {Kaspi}, V.~M., {Pivovaroff}, M.~J., {Kawai},
  N., \& {Tamura}, K. 2002, \apj, 569, 878

\bibitem[{{Gaensler} {et~al.}(2006){Gaensler}, {Chatterjee}, {Slane}, {van der
  Swaluw}, {Camilo}, \& {Hughes}}]{2006ApJ...648.1037G}
{Gaensler}, B.~M., {Chatterjee}, S., {Slane}, P.~O., {van der Swaluw}, E.,
  {Camilo}, F., \& {Hughes}, J.~P. 2006, \apj, 648, 1037

\bibitem[{{Gaensler} \& {Slane}(2006)}]{2006ARA&A..44...17G}
{Gaensler}, B.~M. \& {Slane}, P.~O. 2006, \araa, 44, 17

\bibitem[{{Gallant} \& {Arons}(1994)}]{1994ApJ...435..230G}
{Gallant}, Y.~A. \& {Arons}, J. 1994, \apj, 435, 230

\bibitem[{{Gonzalez} {et~al.}(2006){Gonzalez}, {Kaspi}, {Pivovaroff}, \&
  {Gaensler}}]{2006ApJ...652..569G}
{Gonzalez}, M.~E., {Kaspi}, V.~M., {Pivovaroff}, M.~J., \& {Gaensler}, B.~M.
  2006, \apj, 652, 569

\bibitem[{{Gotthelf}(2004)}]{2004HEAD....8.0807G}
{Gotthelf}, E.~V. 2004, in Bull.\ of the AAS 36, 917

\bibitem[{{Gotthelf} {et~al.}(2002){Gotthelf}, {Halpern}, \&
  {Dodson}}]{2002ApJ...567L.125G}
{Gotthelf}, E.~V., {Halpern}, J.~P., \& {Dodson}, R. 2002, \apjl, 567, L125

\bibitem[{{Gotthelf} {et~al.}(2007){Gotthelf}, {Helfand}, \&
  {Newburgh}}]{2007ApJ...654..267G}
{Gotthelf}, E.~V., {Helfand}, D.~J., \& {Newburgh}, L. 2007, \apj, 654, 267

\bibitem[{Gregory \& Loredo(1992)}]{gl92b}
Gregory, P.~C. \& Loredo, T.~J. 1992, \apj, 398, 146

\bibitem[{{Helfand} {et~al.}(2001){Helfand}, {Gotthelf}, \&
  {Halpern}}]{2001ApJ...556..380H}
{Helfand}, D.~J., {Gotthelf}, E.~V., \& {Halpern}, J.~P. 2001, \apj, 556, 380

\bibitem[{{Hester}(1998)}]{1998MmSAI..69..883H}
{Hester}, J.~J. 1998, Memorie della Societa Astronomica Italiana, 69, 883

\bibitem[{{Kargaltsev} {et~al.}(2007){Kargaltsev}, {Pavlov}, \&
  {Garmire}}]{2007ApJ...660.1413K}
{Kargaltsev}, O., {Pavlov}, G.~G., \& {Garmire}, G.~P. 2007, \apj, 660, 1413

\bibitem[{{Kennel} \& {Coroniti}(1984)}]{1984ApJ...283..694K}
{Kennel}, C.~F. \& {Coroniti}, F.~V. 1984, \apj, 283, 694

\bibitem[{{Kothes} {et~al.}(2008){Kothes}, {Landecker}, {Reich}, {Safi-Harb},
  \& {Arzoumanian}}]{klr+07}
{Kothes}, R., {Landecker}, T.~L., {Reich}, W., {Safi-Harb}, S., \&
  {Arzoumanian}, Z. 2008, \apj, submitted (KLR+08)

\bibitem[{{Landecker} \& {Caswell}(1983)}]{1983AJ.....88.1810L}
{Landecker}, T.~L. \& {Caswell}, J.~L. 1983, \aj, 88, 1810

\bibitem[{{Lu} {et~al.}(2002){Lu}, {Wang}, {Aschenbach}, {Durouchoux}, \&
  {Song}}]{2002ApJ...568L..49L}
{Lu}, F.~J., {Wang}, Q.~D., {Aschenbach}, B., {Durouchoux}, P., \& {Song},
  L.~M. 2002, \apjl, 568, L49

\bibitem[{{Lyubarsky}(2002)}]{2002MNRAS.329L..34L}
{Lyubarsky}, Y.~E. 2002, \mnras, 329, L34

\bibitem[{{Mori} {et~al.}(2001){Mori}, {Tsunemi}, {Miyata}, {Baluta},
  {Burrows}, {Garmire}, \& {Chartas}}]{2001ASPC..251..576M}
{Mori}, K., {Tsunemi}, H., {Miyata}, E., {Baluta}, C.~J., {Burrows}, D.~N.,
  {Garmire}, G.~P., \& {Chartas}, G. 2001, in ASP Conf. Ser. 251: New Century
  of X-ray Astronomy, ed. H.~{Inoue} \& H.~{Kunieda}, 576

\bibitem[{{Ng} \& {Romani}(2004)}]{2004ApJ...601..479N}
{Ng}, C.-Y. \& {Romani}, R.~W. 2004, \apj, 601, 479

\bibitem[{{Pavlov} {et~al.}(2001){Pavlov}, {Kargaltsev}, {Sanwal}, \&
  {Garmire}}]{2001ApJ...554L.189P}
{Pavlov}, G.~G., {Kargaltsev}, O.~Y., {Sanwal}, D., \& {Garmire}, G.~P. 2001,
  \apjl, 554, L189

\bibitem[{{Pavlov} {et~al.}(1995){Pavlov}, {Shibanov}, {Zavlin}, \&
  {Meyer}}]{1995lns..conf...71P}
{Pavlov}, G.~G., {Shibanov}, Y.~A., {Zavlin}, V.~E., \& {Meyer}, R.~D. 1995, in
  Proc.\ of the NATO Adv.\ Study Inst.\ on the Lives of the Neutron Stars 3,
  ed.\ M.~A. Alpar, U. Kiziloglu, J. van Paradijs (Dordrecht: Kluwer), 71

\bibitem[{{Petre} {et~al.}(2007){Petre}, {Hwang}, {Holt}, {Safi-Harb}, \&
  {Williams}}]{2007ApJ...662..988P}
{Petre}, R., {Hwang}, U., {Holt}, S.~S., {Safi-Harb}, S., \& {Williams}, R.~M.
  2007, \apj, 662, 988

\bibitem[{{Sefako} \& {de Jager}(2003)}]{2003ApJ...593.1013S}
{Sefako}, R.~R. \& {de Jager}, O.~C. 2003, \apj, 593, 1013

\bibitem[{{Slane} {et~al.}(2004){Slane}, {Helfand}, {van der Swaluw}, \&
  {Murray}}]{2004ApJ...616..403S}
{Slane}, P., {Helfand}, D.~J., {van der Swaluw}, E., \& {Murray}, S.~S. 2004,
  \apj, 616, 403

\bibitem[{{Stappers} {et~al.}(2003){Stappers}, {Gaensler}, {Kaspi}, {van der
  Klis}, \& {Lewin}}]{2003Sci...299.1372S}
{Stappers}, B.~W., {Gaensler}, B.~M., {Kaspi}, V.~M., {van der Klis}, M., \&
  {Lewin}, W.~H.~G. 2003, Science, 299, 1372

\bibitem[{{Tsuruta} {et~al.}(2002){Tsuruta}, {Teter}, {Takatsuka}, {Tatsumi},
  \& {Tamagaki}}]{2002ApJ...571L.143T}
{Tsuruta}, S., {Teter}, M.~A., {Takatsuka}, T., {Tatsumi}, T., \& {Tamagaki},
  R. 2002, \apjl, 571, L143

\bibitem[{{van der Swaluw} {et~al.}(2001){van der Swaluw}, {Achterberg},
  {Gallant}, \& {T{\'o}th}}]{2001A&A...380..309V}
{van der Swaluw}, E., {Achterberg}, A., {Gallant}, Y.~A., \& {T{\'o}th}, G.
  2001, \aap, 380, 309

\bibitem[{{Velusamy} {et~al.}(1989){Velusamy}, {Becker}, {Goss}, \&
  {Helfand}}]{1989JApA...10..161V}
{Velusamy}, T., {Becker}, R.~H., {Goss}, W.~M., \& {Helfand}, D.~J. 1989,
  Journal of Astrophysics and Astronomy, 10, 161

\bibitem[{{Weisskopf} {et~al.}(2000){Weisskopf}, {Hester}, {Tennant}, {Elsner},
  {Schulz}, {Marshall}, {Karovska}, {Nichols}, {Swartz}, {Kolodziejczak}, \&
  {O'Dell}}]{2000ApJ...536L..81W}
{Weisskopf}, M.~C., 
  et~al.\
  2000, \apjl, 536, L81

\bibitem[{{Zavlin} {et~al.}(1996){Zavlin}, {Pavlov}, \&
  {Shibanov}}]{1996A&A...315..141Z}
{Zavlin}, V.~E., {Pavlov}, G.~G., \& {Shibanov}, Y.~A. 1996, \aap, 315, 141

\end{thebibliography}

\clearpage

\begin{deluxetable}{@{}cccccccc}

\tablecaption{\label{tab:spec}Spectral fitting results for X-ray features in DA 495}

\tablenum{1}

\tablehead{\colhead{Emission} & \colhead{Model} & \colhead{$\nh$} & \colhead{$\log(T_{\rm eff})$} & \colhead{$\Gamma$} & \colhead{$\chi^2/\nu$} & \colhead{Dist.} & \colhead{$\log(L_X)$} \\ 
\colhead{feature} & \colhead{} & \colhead{($10^{21}$ cm$^{-2}$)} & \colhead{(K)} & \colhead{} & \colhead{} & \colhead{(kpc)} & \colhead{(ergs s$^{-1}$)} } 

\startdata
Pt. source & BB & $2.7\pm2.0$ & $6.40\pm0.10$ & \ldots & 4.4/7 & \ldots & 31.0 \\
 & H (0 G) 	   & $6.4\pm0.7$ & $\sim5.8$ & \ldots & 5.2/7 & 0.8 & 31.7 \\
 & H ($10^{12}$ G) & $6.9\pm0.9$ & $\sim6.0$ & \ldots & 5.9/7 & 1.5 & 32.3 \\
 & H ($10^{13}$ G) & $6.8\pm0.9$ & $\sim6.0$ & \ldots & 5.1/7 & 1.4 & 32.3 \\
\hline
Nebula & PL & $1.9\pm1.7$ & \ldots & $1.58\pm0.32$ & 8.4/19 & \ldots & 31.4 \\
\hline
Pt. source  & BB & $2.3\pm1.7$ & $6.42\pm0.08$ & \ldots & 12.9/27 & \ldots & 31.0 \\
+ nebula & + PL &                & \ldots   & $1.63\pm0.27$ &  & \ldots & 31.3 \\
\hline
Pt. source  & H ($10^{13}$ G) & $6.0\pm0.7$ & $\sim 6.0$ & \ldots & 18.7/27 & 1.8 & 32.4 \\
+ nebula & + PL & & \ldots & $2.10\pm0.24$ & & \ldots & 31.8 \\
\enddata

\tablecomments{Models BB and PL refer to blackbody and power-law spectra,
respectively. H refers to the hydrogen-atmosphere NSA models 
with magnetic field strengths $B$ given in parentheses. Luminosities, which 
for thermal and nonthermal models are bolometric and in the 2--10 keV band, 
respectively, assume a distance of 1 kpc if the spectral model does not provide
an independent estimate.}

\end{deluxetable}

\clearpage

\end{document}